# Theoretical Study of Extensive Air Shower Effects in Atmosphere by Simulating the Lateral structure of Several Cosmic Radiations

Hassanen Abdulhussaen Jassim, A. A. Al-Rubaiee*, Iman Tarik Al-Alawy

College of Science, Department of Physics, Al-Mustansiriyah University, Baghdad, Iraq

*dr.rubaiee@uomustansiriyah.edu.iq

**Abstract**

Extensive air showers (EAS) are a cascade of electromagnetic radiation and ionized particles that produced in the atmosphere through the interaction of a primary cosmic ray with the atom's nucleus in the air producing a huge amount of secondary particles such as X-ray, electrons, neutrons, muons, alpha particles, etc. In this work, EAS effects were demonstrated by estimating the lateral distribution function (LDF) at ultrahigh energies of the various cosmic ray particles. The LDF of charged particles such as electron and positron pair production, gamma and muons particles was simulated at ultrahigh energies $10^{16}$, $10^{18}$ and $10^{19}$ eV. The simulation was carried out using an air shower simulator called AIRES system version 2.6.0. The effect of the primary particles, energies and zenith angle (θ) on the LDF of charged particles produced in the EAS was taken into account. Comparison of LDF for charged particles and experimental results gave good agreement for electron and positron pair production and muons particles at $10^{19}$ eV for θ=0° and 10°.

## 1. Introduction

High energy cosmic rays have been detected via the EAS that produced in the Earth's atmosphere [1]. In 1930, Pierre Victor Auger, French physicist was discovered EAS through producing more and more particles in the atmosphere [2]. A little part of the primary particle's kinetic energy has been converted into mass energy. Then, the residual kinetic energy can be distributed through the shower. The multiplication process is still continuing till the EAS particles energy still insufficient for producing more and more particles in sequential collisions. This stage of shower development can be called the maximum of shower [2, 3]. The LDF of charged particle in EAS is a requisite quantity for the observations of the Earth's cosmic radiation, which are mostly derived from the EAS observables [4]. The parameter that utilized for describing the shape of the lateral structure density is the lateral shape parameter in the NKG function "Nishimura-Kamata-Greisen function" [5, 6]. The EAS develops in a convoluted way as a combination of electromagnetic and hadronic showers. It is important to fulfill a detailed numerical simulation of the EAS for inferring the primary comic radiation properties, which produce them. The number of charged particles in ultra high energy EAS may be enormous and may exceeding $10^{10}$, therefore the simulations are an important way to challenge this situation [7]. Before the era of high-speed computing, Heitler presented a very simple model for the development of electromagnetic cascade [8]. At the time, Heitler, Rossi, and Gaisser developed more sophisticated analytical tools, which included more physical influences [9, 10]. In 2008, Cotzomi studied some observations about the LDF of charged particles at energy above $10^{17}$ eV

[11]. In 2013, the age parameter of the lateral structure has been studied by Tapia by estimating the chemical composition of EAS particles [12]. Recently, (in 2018), Ivanov was studied the distribution of the zenith angle of the cosmic ray showers measured with the Yakutsk array and its application to the analysis of arrival directions in equatorial coordinates [13].

The results of the current calculations show that the density of charged particles reaches the Earth's surface, such as the electron and positron pair production, gamma and muons particles, by simulating the LDF performed using the Monte Carlo AIRES system at ultrahigh energies $10^{16}$, $10^{18}$ and $10^{19}$eV. The comparison of the estimated LDF of the charged particles such as the electron and positron pair production and muons with the simulated results by Sciutto and Yakutsk EAS observatory gives good agreement at $10^{19}$eV [14, 15].

## 2. Electromagnetic and Hadronic Showers

Heitler's model is the simplest conception of electromagnetic cascades and can extends it to the EAS. The purpose of using a very simple model is to show clearly the physics involved. Nevertheless, the most significant properties of the electromagnetic cascades has been predicted by Heitler's model [7]. In the Figure 1a was shown a single photon that radiated by a single electron after transmitting $d$ splitting length [7]:

$$d = \lambda_r \ln 2 \quad \text{.......} \quad (1)$$

Where, $\lambda_r$ is the radiation length in the medium; $d$ is the distance over which an electron loses, on average, half of its energy by radiation. After traveling the same distance, the photon is split into $e^\pm$ pair. In either case, the particle's energy (electron or photon) is assumed to be evenly divided between two couple particles. After $n$ splitting lengths, the distance $x$ is given by:

$$x = n \lambda_r \ln 2 \quad \text{.......} \quad (2)$$

The total size of the shower for the electrons and photons is:

$$N = 2^n = e^{x/\lambda_r} \quad \text{.......} \quad (3)$$

When the particle energies become too low for pair production, the multiplication of EAS particles ceases. Therefore, Heitler takes this energy of electron as a critical energy that is given by the symbol ($\xi_c^e$). The average energy lose of collision will exceed the radiative losses.

Through supposing a shower that initiated by a single primary photon with $E_o$ primary energy, the cascade in this shower will reach its maximum size ( $N = N_{max}$) when the produced particles have a critical energy $\xi_c^e$, therefore [7]:

$$E_\circ = \xi_c^e N_{max} \quad \text{.......} \quad (4)$$

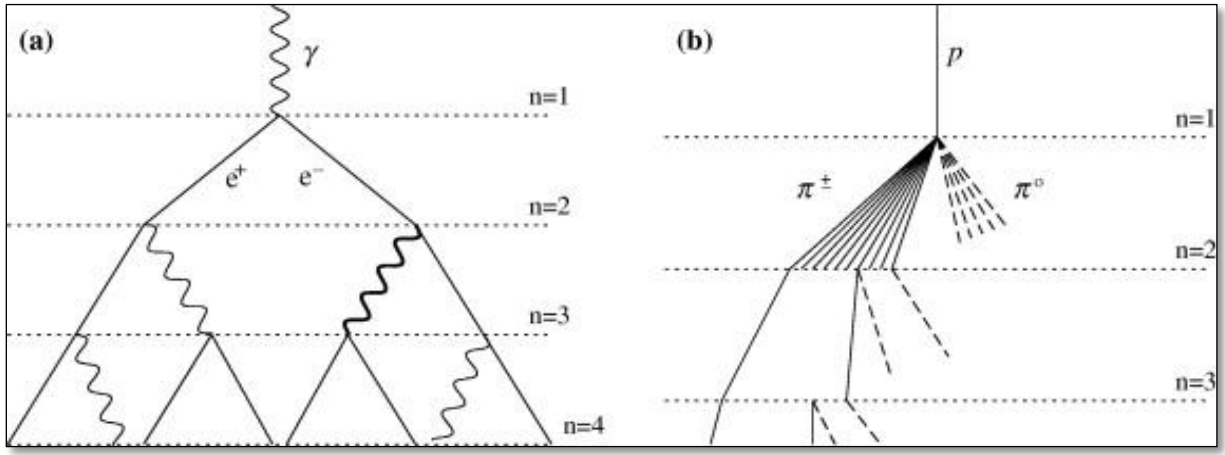

**Figure 1:** Schematic views of (a) Electromagnetic cascade and (b) Hadronic shower[7].

As shown in figure 1b, the air showers initiated by the hadrons were designed using a method similar to Heitler's method [7]. Charged ($\pi^{\pm}$) and neutral ($\pi^o$) pions produced when hadron particles traversing one layer in atmosphere. Through this process, the $\pi^o$ pions directly will be decayed to photon particles and therefore the electromagnetic cascade initiated [7]. Figure 1b demonstrates the electromagnetic showers through the decaying of the $\pi^o$ pions (dashed lines). While the $\pi^{\pm}$ pions will continue interacting in the shower until their energy become below the critical pion energy ($\xi_c^{\pi}$). Accordingly, the $\pi^{\pm}$ pions will decay to muons that reach the earth [7].

## 3. Lateral Distribution Function

The LDF of charged particles in EAS is an important quantity of the ground surveillance for cosmic radiations, from which most cascade observables are deduced [12]. Studying of EAS may be performed experimentally on the surface of the ground, underground and at many mountain rising through determining some LDF quantities, i.e. the density of charged particles initiated in EAS as a function of the shower core distance or in another meaning, the LDF is the shower structure of the cascade at various depths in atmosphere [2]. The expression that used extensively for describing the LDF form is the NKG function that is given through the forum [5]:

$$\rho(r) = \frac{N_e}{2*\pi*R_M^2} * C(s) * (\frac{r}{R_M})^{(s-2)} * (\frac{r}{R_M} + 1)^{(s-4.5)} \ldots\ldots (4)$$

Where $\rho(r)$ is the particle density on the distance $r$ from the shower core, $N_e$ is the total number of shower electrons, $R_M$ =118 m is Molier radii, $s$ is the shower age parameter, and $C(s)$ is the normalizing factor which is equal $0.366\, s^2 * (2.07 - s)^{1.25}$ [16].

## 4. Results and Discussion
### 4.1 Simulating of LDF using AIRES system

AIRES is an acronym for AIR-shower Extended Simulations, which is defined as a set of programs and subroutines that used for simulating the EAS particle cascades, which initiated after interaction of the primary cosmic radiations with a high energies in the atmosphere and manage all the output associated data [14]. AIRES provides full space-time particle propagation in a true medium, where the features of the atmosphere, the geomagnetic field and Earth's curvature are taken into account adequately[14].

There are many particles are taken into account through the simulation using AIRES system such as: "electrons; positrons; gammas and muons". The incident primary particle in EAS may be primary proton or iron nuclei or other primaries that mentioned in the AIRES guide document with a very high primary energies that may exceeding $10^{21}$ eV [14].

Figure 2 shows the density of several secondary particles as a function of the distance from the shower axis that reaches the Earth's surface by AIRES simulation. The effect of the primary particles (proton and iron) and energies ($10^{16}$, $10^{18}$ and $10^{19}$eV) on the density of charged particles produced in the EAS was taken into consideration. As shown in figure 2, the density of several secondary particles decreases with increasing the distance from the shower axis. Finally, the primary energy is divided between muons and electromagnetic particles (electron and positron pair production and gamma particles) in sub-showers.

Figure 3 displays the effect of the primary particle energies on the density of secondary particles for zenith angle ($\theta = 0°$, $10°$ and $30°$) of the primary particles (proton and iron) by AIRES simulation. Through this figure one can see that the lateral density of different secondary particles is directly proportion to the primary particle energy, i.e. the lateral density increases with increasing the primary energy.

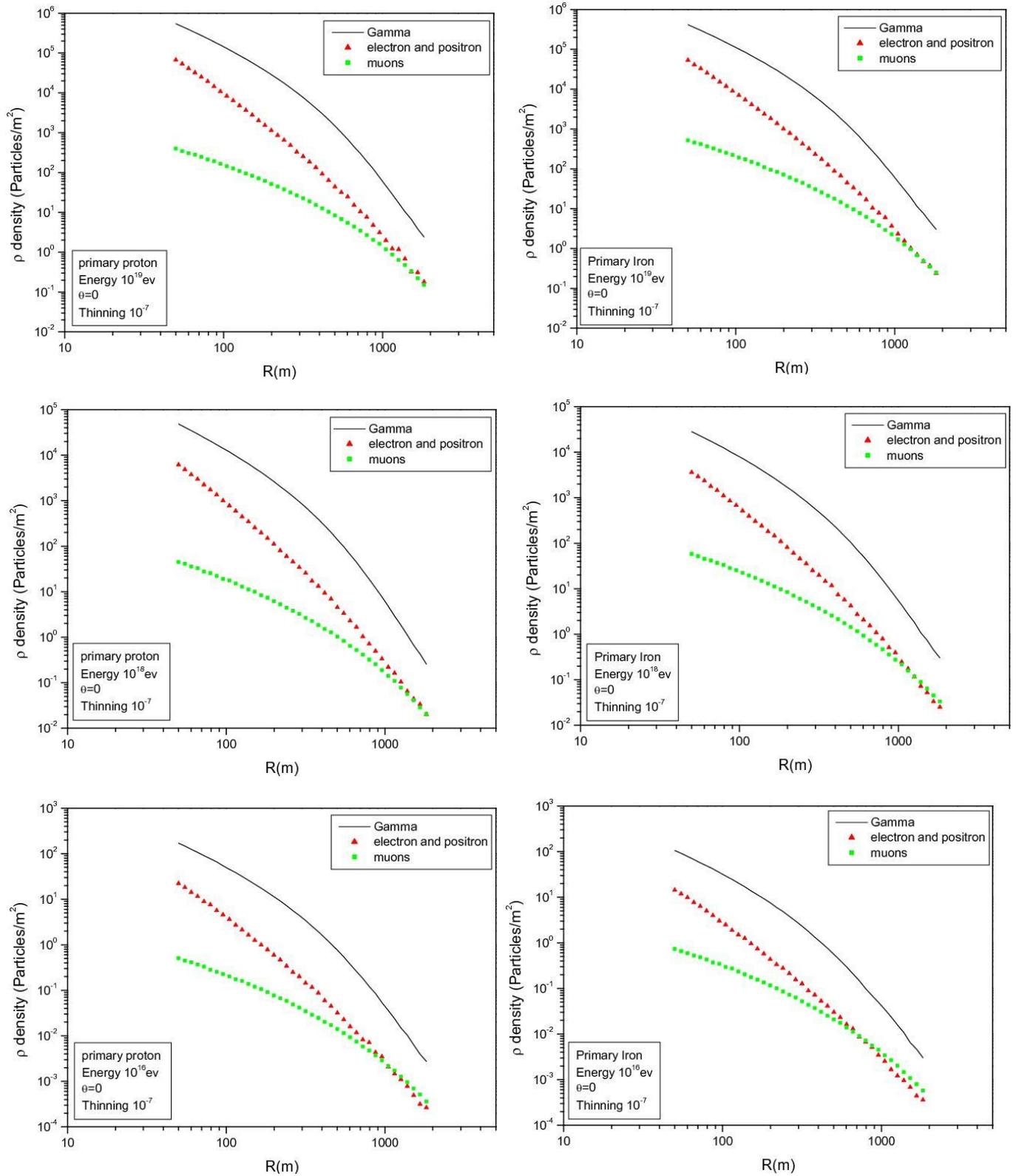

**Figure 2:** Densities of several secondary particles as a function of *R* that reaches the Earth's surface for vertical showers at energies ($10^{16}$, $10^{18}$ and $10^{19}$ eV) of p and Fe.

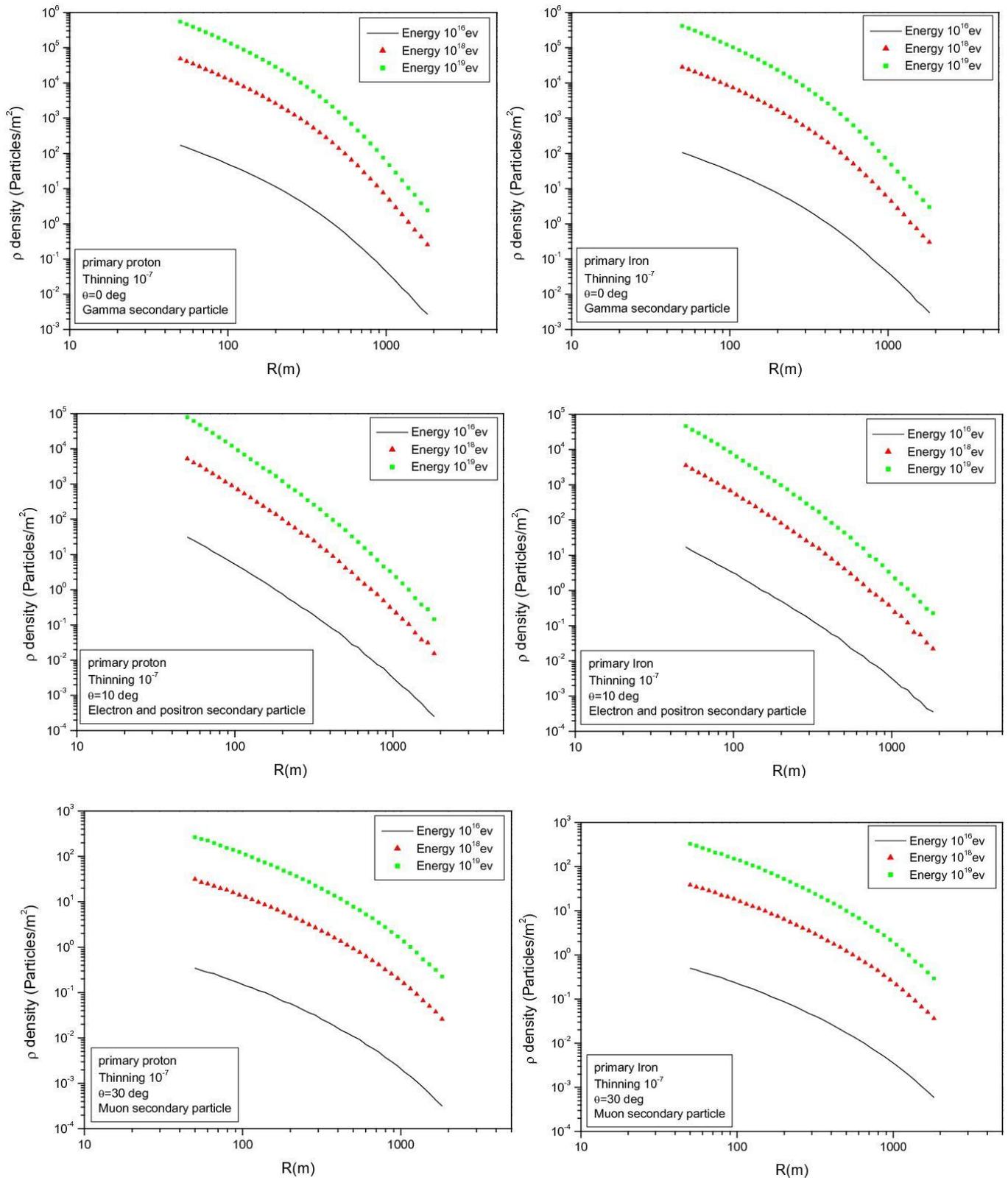

**Figure 3: The primary energy effects on secondary particle densities for primary p and Fe at different zenith angles (θ = 0°, 10° and 30°).**

## 4.2 Comparison with Sciutto experience and the Yakutsk Observatory

Figure 4 demonstrates the comparison between the present results of LDF that performed by AIRES simulation (solid lines) with the results simulated by Sciutto (triangle symbols)[14].This figure displayed a good agreement of the secondary particles (electron and positron) and muons particles that were initiated by primary proton at energy $10^{19}$eV and vertical EAS showers.

The Yakutsk EAS array studies the very high energy cosmic radiations, which occurs in the field of astrophysics, that is, an important area in the physics. There are two main goals for the construction of the Yakutsk EAS observatory; the $1^{st}$ goal is the elementary particles investigating of the cascades that initiated by the primary particles in atmosphere. The $2^{nd}$ goal is the astrophysical characteristics reconstruction of the primary particles such as: " mass composition, energy spectrum, intensity and the their origin" [15].

Figure 5 shows the comparison between the present results with the experimental data that obtained by Yakutsk Observatory [15].The curves in this figure displayed a good agreement for(electron and positron) and muons particles that initiated by primary proton at energy $10^{19}$ eV and slanted EAS showers with $\theta = 10°$.

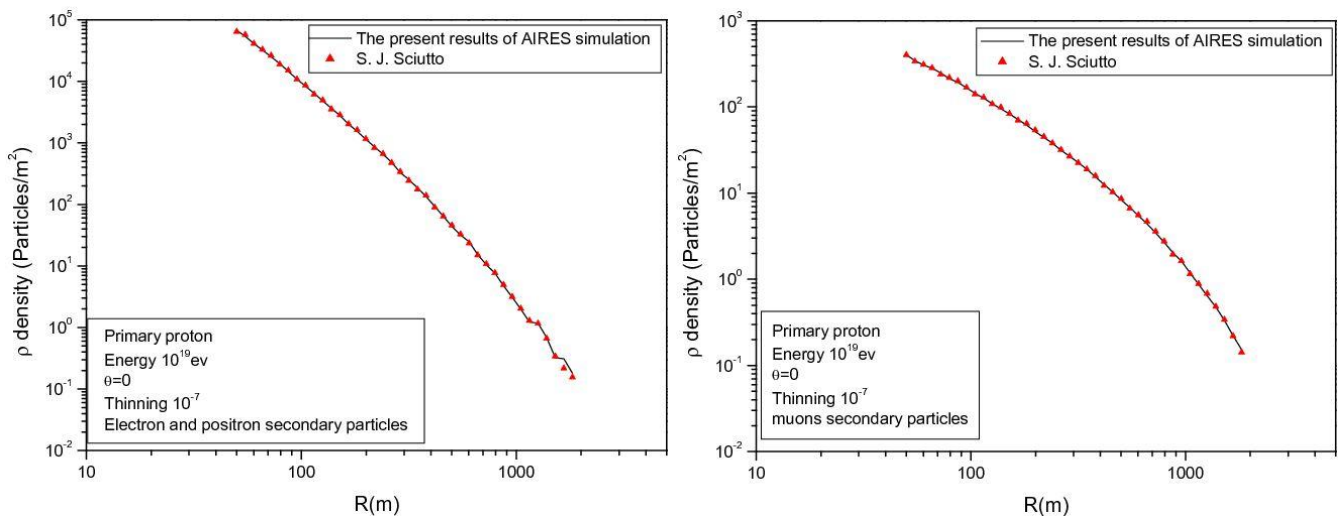

**Figure 4: Comparison between the present results of simulated LDF by AIRES system with the results that simulated by Sciutto for primary p at energy$10^{19}$ eVfor (electron and positron) and muons secondary particles.**

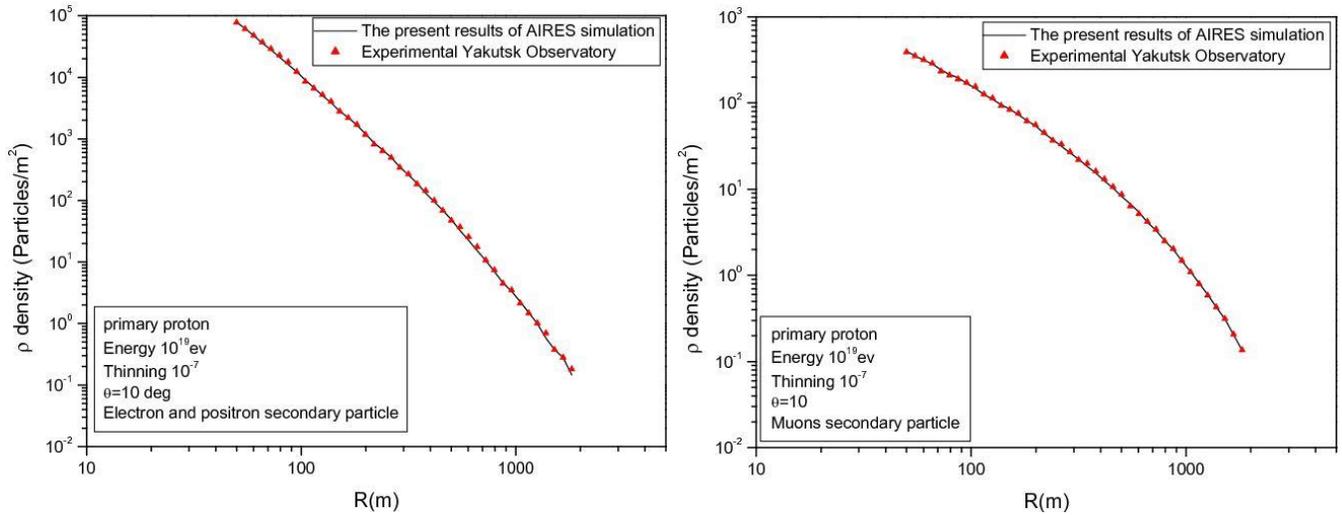

**Figure 5:** Comparison between the present results of simulated LDF by AIRES system with the experimental data obtained by Yakutsk Observatory for primary p at energy $10^{19}$ eV for (electron and positron) and muons secondary particles.

**5.  Conclusions**

In the present work, the lateral distribution function of charged particles using AIRES system for two primary particles such as (proton and iron nuclei) was simulated at different ultrahigh energies $10^{16}$, $10^{18}$ and $10^{19}$ eV. The simulation of charged particle lateral structure demonstrates ability for identifying the primary cosmic ray particle and its energy. The important advantage of the present work is to make a library of Lateral structure samples that may utilized for analyzing real EAS events that detected and registered with an EAS arrays.

The introduced results using AIRES system are identified with Yakutsk experimental data, proving that AIRES system provides an appropriate environment for the study of high-energy cosmic rays. Therefore, charged particles reaching the Earth's surface have many effects on weather, human health and other effects.